\documentclass{webofc}
\usepackage[varg]{txfonts}   
\usepackage{hyperref}
\usepackage[mathlines]{lineno}
\begin{document}
\title{Coffea-Casa: Building composable analysis facilities for the HL-LHC}

\author{\firstname{Sam} \lastname{Albin}\inst{1}
\and
\firstname{Garhan} \lastname{Attebury}\inst{1}
\and
\firstname{Kenneth} \lastname{Bloom}\inst{1}
\and
\firstname{Brian} \lastname{Bockelman}\inst{2}
\and
\firstname{Carl} \lastname{Lundstedt}\inst{1}
\and
\firstname{Oksana} \lastname{Shadura}\inst{1}
\fnsep\thanks{\email{oksana.shadura@cern.ch}}
\and
\firstname{John} \lastname{Thiltges}\inst{1}
}

\institute{University of Nebraska-Lincoln, Lincoln, NE 68588 
\and
          Morgridge Institute for Research, 330 N. Orchard Street, Madison, WI 53715
          }

\abstract{%
The large data volumes expected from the High Luminosity LHC (HL-LHC) present challenges to existing paradigms and facilities for end-user data analysis. Modern cyberinfrastructure tools provide a diverse set of services that can be composed into a system that provides physicists with powerful tools that give them straightforward access to large computing resources, with low barriers to entry. The Coffea-Casa analysis facility (AF) provides an environment for end users enabling the execution of increasingly complex analyses such as those demonstrated by the Analysis Grand Challenge (AGC) and capturing the features that physicists will need for the HL-LHC.

We describe the development progress of the Coffea-Casa facility featuring its modularity while demonstrating the ability to port and customize the facility software stack to other locations. The facility also facilitates the support of batch systems while staying Kubernetes-native. We present the evolved architecture of the facility, such as the integration of advanced data delivery services (e.g. ServiceX) and making data caching services (e.g. XCache) available to end users of the facility. We also highlight the composability of modern cyberinfrastructure tools. To enable machine learning pipelines at coffee-casa analysis facilities, a set of industry ML solutions adopted for HEP columnar analysis were integrated on top of existing facility services. These services also feature transparent access for user workflows to GPUs available at a facility via inference servers while using Kubernetes as enabling technology.
}
\maketitle
\section{Introduction}
\label{intro}
 
\begin{figure}
\centering
\sidecaption
\includegraphics[width=0.7\textwidth]{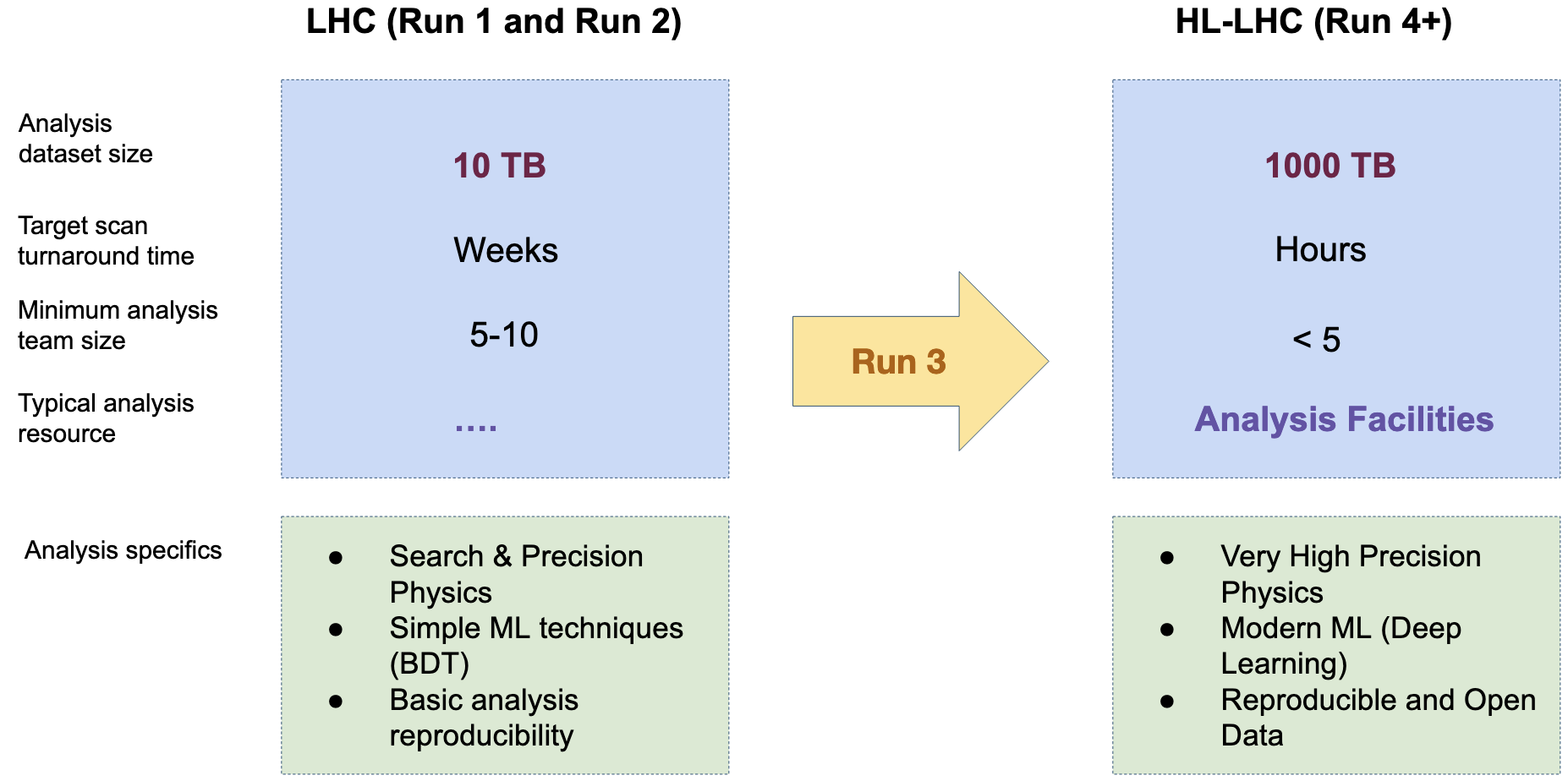}
\caption{Requirements for the physics analysis and computing infrastructure for HL-LHC}
\label{fig-1}       
\end{figure}

The HL-LHC will be a challenging analysis environment compared to the LHC experiments of today. The data volumes will go up by a factor 100 and and to achieve the desired physics reach of results, analysts will need new techniques and approaches as well new software infrastructure, see figure \ref{fig-1}. An analysis that physicists are able to do on a laptop today by simply having files with physics events stored on local disk would  be expected to be handled on a dedicated facility during the HL-LHC era, leveraging complex large-scale computing hardware, advanced data delivery services, and ML training and inference. 

Existing facilities  provide the back-bone setups for physicists doing analysis, but can be complex for users who need to track different configurations, different ways to access data, or trying to easily move from one facility to other. To manage the different interfaces and different scaling mechanisms on various facilities, together with the lack of documentation, could be challenging experience especially for new users. Also, not all existing facilities are suitable for interactive analysis use. 

The Coffea-Casa analysis facility \cite{adamec2021coffea} starts with a base of the “Coffea” processing framework \cite{smith_nick_2020} for low-latency columnar analysis and uses a modular approach, adding other services such as ServiceX \cite{servicex} or a simple scale-out of tasks into a traditional batch system. The facility provides an interactive experience for physicists that is closer to working on a laptop as opposed to a traditional batch system-based facility. 

The facility adopts an approach that allows transforming existing computing facilities into composable systems using Kubernetes as the enabling technology.

In this article we will describe the detailed overview of design approaches behind the Coffea-Casa analysis facility. We will also describe the experience of using the facility as a testbed for early adopters to investigate the Python analysis ecosystem and additional services. In this way we hope to reach the goal of handling HL-LHC analysis requirements and expected data rates.
 
\section{Building blocks for Coffea-Casa analysis facility}

Over the last few years, the Coffea-Casa analysis facility development has validated essential design features, enumerated in figure 2, for use in future facilities.

Coffea-Casa's basic framework is an extension of the Zero-to-JupyterHub project with custom container images, services and classes containing extended functionality. This creates a basic framework on which to build custom functionality to satisfy the design features of Coffea-Casa.  Columnar analysis and the Pythonic ecosystem is supported in the custom containers spun up in the Kubernetes~\cite{brewer2015kubernetes} instance of Coffea-Casa.  Efficient data delivery, data management and data caching solutions are facilitated using an XCache\cite{XCache} service. Machine Learning (ML) services and tools are provided by Kubernetes based MLFlow \cite{mlflow} and Triton Inference services \cite{triton}. Custom modified Dask-Jobqueue \cite{dask-jq} classes allow for scalable integration with large scale compute resources. All of these are managed via modern CI/CD techniques. These modules will be outlined in more detail in the rest of this article.

\begin{figure}
\centering
\sidecaption
\includegraphics[width=0.7\textwidth]{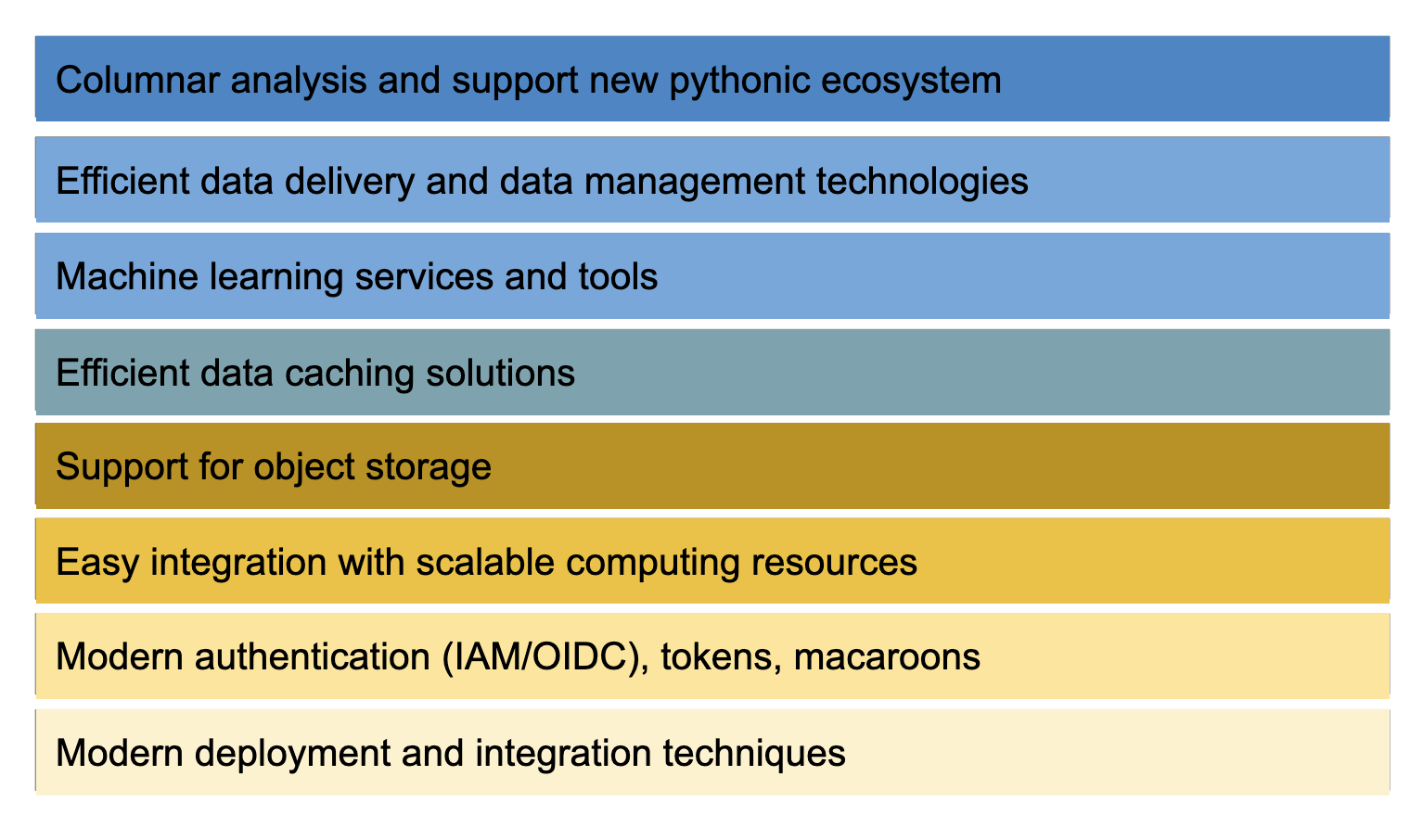}
\caption{Essential design features for future analysis facility}
\label{fig-2}       
\end{figure}

\subsection{Easy integration with scalable computing resources
}
One of the key design concepts in Coffea-Casa was to leverage currently deployed compute and storage infrastructure to facilitate analysis work. Experiments have invested heavily in the creation and deployment of compute infrastructure designed to serve current workflow paradigms. In order for Coffea-Casa to be readily adopted, it must be able to utilize this infrastructure without modifying the existing infrastructure.  

\begin{figure}
\centering
\sidecaption
\includegraphics[width=0.7\textwidth]{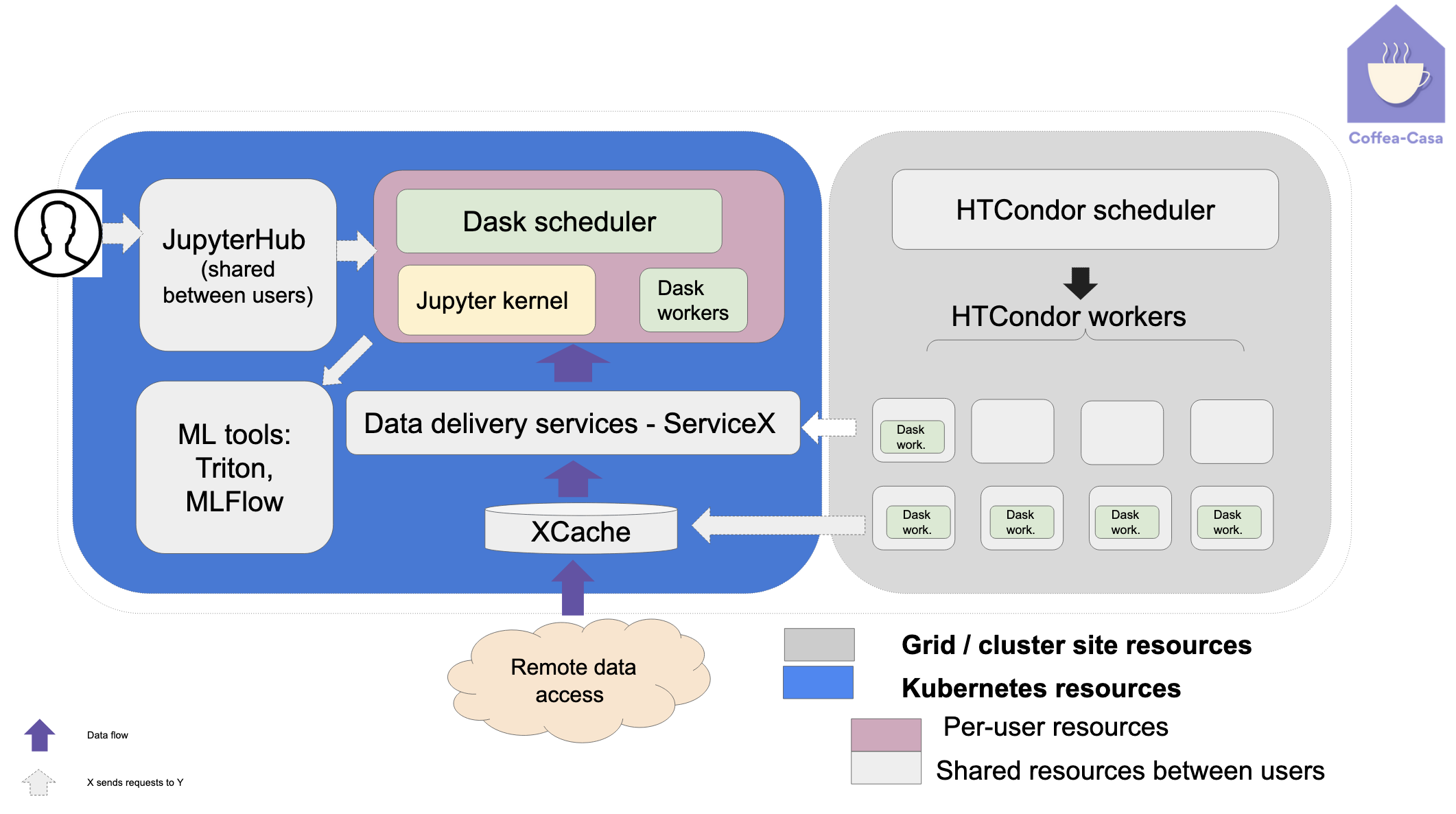}
\caption{The schema of Coffea-Casa Analysis Facility}
\label{fig-3}       
\end{figure}

To facilitate this design imperative, Coffea-Casa was created to be easily integrated with existing batch resources.  Figure \ref{fig-3} shows how a Coffea-Casa instance integrates with an HTCondor-based resource.  Here the HTCondor \cite{condor-practice} resource is entirely external to the Coffea-Casa instance.  As the Coffea-Casa workload scales up, it submits Dask \cite{dask} jobs to the HTCondor queue.  These jobs start a container with a Dask worker service that connects back to the Dask scheduler created for the user's instance when the user logged in.  In this way the HTCondor resource can be easily utilized to do Dask work for the Coffea-Casa user. 

The integration with other types of batch resources is readily accomplished by the modular nature of Coffea-Casa.

\subsection{Transition to tokens}

In past analysis workflows, authentication and authorization was handled through the use of x509 proxy certificates and complex mapping mechanisms.  This use of proxy certificates has a couple of inherent draw backs.  First, x509 proxies are generally created with a long lifetime to survive for the entire analysis workflow.  Second, there is no capabilities associated with a given proxy; the bearer of the proxy is given access and authorization based solely on identity and not on desired functionality.   There does not exist any way for an x509 proxy to have its authorized capabilities stored within the proxy itself. The transition to token based authentication seeks to remedy these two shortcomings.

First, token lifetimes are created with a much shorter valid lifetime and thus present a smaller window of opportunity for being compromised.  A token is able to be renewed to allow for authentication and authorization to persist through the lifetime of the analysis workflow while any individual token remains short lived.

Second, tokens can be constructed to have only specific permissions and capabilities, allowing for fine-grained authorization without the need for complex mapping infrastructure.  For instance, a token may be constructed to have read privileges for a storage element but lack any write privileges, thus creating a more secure workflow.

Coffea-Casa seeks to remove, where it can, any use of x509 proxies by the user and instead facilitate authentication and authorization via tokens.  These tokens are constructed upon user login based on their identity returned by the OpenID Connect (OIDC) layer \cite{sakimura2014openid} of an OAuth service to which the Coffea-Casa instance is registered.  In this way a user's identity is verified once at login and the user need not handle identity management manually.

Numerous tokens are constructed and managed by Coffea-Casa including tokens to allow reading from an XCache \cite{bauerdick2014xrootd} instance and tokens for submitting to an HTCondor queue. 

\subsection{Integrating data delivery services and advanced caching}

The XRootD software framework is foundational for efforts across the LHC. It is used as a reference platform in IRIS-HEP \cite{iris-hep} for data streaming and bulk data transfers, as the basis for data federations in CMS, and, in its “XCache” configuration, as a data caching service for both production and analysis. All instances of Coffea-Casa have integrated XCache support  for experimental datasets.  This allows for data access speedup during the interactive analysis in cases where the user needs to access the dataset multiple times.

For columnar analysis, which is one of the growing approaches to physics analysis, the ServiceX service was developed to run inside analysis facilities.  ServiceX is designed to create and cache columns to meet an analyst's data delivery needs.
ServiceX is easily deployed in Kubernetes as a part of facility deployment, see figure \ref{fig-3}, and available for each user directly from their session through pre-generated configuration files and tokens.

\subsection{Machine learning services and tools for analysis facilities}

There are many stand-alone tools to aid with building, training, optimization, and deployment of machine learning (ML) models (e.g. MLFlow or KubeFlow \cite{bisong2019kubeflow}), along with more traditional approaches commonly used in the community. 

When moving from running individual machine learning workflows on a laptop to running on facilities able to run them efficiently on scale,  ML Operations (MLOps) \cite{alla2021mlops} can help. It is a methodology for enabling collaboration across multiple scientists. MLOps helps to gain control over different model versions, multiple experiments within the same problem, and model management and deployment. Adopting MLOps in the analysis workflows allows admins to automate all the steps and incorporate CI/CD practices.

Providing MLOps infrastructure requires expertise to deploy and provide relevant MLOps software infrastructure and services. While technically challenging in some cases, it contributes a significant added value for turnaround times on physics analyses that rely heavily on ML approaches.

There are multiple industry solutions for various use cases, but as a part of the Analysis Grand Challenge analysis pipeline \cite{agc_rtd}, the main focus is to provide a platform for handling the ML life-cycle and ML inference server often used in HEP analysis.

Two services, MLFlow and Nvidia Triton were deployed and integrated in the Coffea-Casa  analysis facility, again see figure \ref{fig-3}, and tested using a typical columnar data analysis workflow. MLflow is an open-source platform that provides experiment and model run management at the core of their platform and can be easily deployed via Kubernetes. NVIDIA’s Triton Inference server provides the transparent access to high-speed, GPU-based inference as part of a user's application.

\subsection{Connecting all together: Coffea-Casa development strategies}

While developing Coffea-Casa we are relying on a GitOps strategy \cite{beetz2021gitops}. GitOps is defined as a model for operating Kubernetes clusters or cloud-native applications (e.g. Coffea-Casa AF). The main advantage of it is that it implements the “infrastructure-as-a-code” concept. It allows for rapid collaboration, better quality control, and higher level of automation (CD/CI).

Using GitOps for the analysis facility development allowed us to easily handle configuration of the facility via a collaborative group of administrators in a deterministic manner. It also allows us to easily package the core infrastructure as a Helm chart for redeployment in other facilities.

\section{IRIS-HEP Analysis Grand Challenge}

The IRIS-HEP Analysis Grand Challenge (AGC) \cite{held_alexander_2022} started out as an integration exercise for IRIS-HEP to provide mechanisms for testing an end-to-end analysis pipeline. It is designed to prepare infrastructure  for the HL-LHC in the context of a physics analysis of realistic scope and scale and to develop flexible, easy-to-use, low latency analysis facilities.  In addition, it allows evaluation of the new Python ecosystem for analysis. 

There are two components to the AGC \cite{agc_code} for this purpose: the definition of a physics analysis task representative of HL-LHC requirements and the implementation of an analysis pipeline addressing this task.

The AGC provides a well-defined physics analysis task with a pipeline implementation. It allows scientists to identify and address performance bottlenecks and usability issues. There have been multiple AGC implementations developed, which use coffea \cite{coffea} (the IRIS-HEP implementation), RDataframe \cite{piparo2019rdataframe}, Julia \cite{bezanson2017julia}, and columnflow \cite{cf}. 

The goal of the AGC is to bridge analysis at scale gap towards HL-LHC and provide closer connections to LHC experiments.  An additional goal is providing extended functionality and testing data preservation pipelines on the Coffea-Casa facility. We also plan to stress test our facility with a series of AGC benchmarks with incremental data rate goals for throughput.

\section{Conclusions}

Coffea-Casa is a prototype analysis facility delivering extra functionality needed for improved user experience. It allows us to rethink established design patterns and integrate new advanced services with traditional facilities enabling the possibility of quick interactive analysis turnaround, allowing end-users to worry only about physics. 

We believe focusing on enabling ML-based analysis for facilities together with the ability to handle HL-LHC data volumes is the right path to future analysis facilities.

\section{Acknowledgements}
This work was supported by the U.S. National Science Foundation (NSF) Cooperative Agreement OAC-1836650 (IRIS-HEP) and NSF-2121686.

\bibliography{bib/references}

\end{document}